\documentclass[]{aastex62}
\usepackage{CJKutf8}
\usepackage{comment}
\usepackage{amsmath}
\usepackage{color}
\usepackage{natbib}
\usepackage{amssymb}
\usepackage{amsfonts}
\usepackage{footmisc}

\newcommand{\beq}{\begin{equation}}
\newcommand{\eeq}{\end{equation}}

\begin{document}
\begin{CJK*}{UTF8}{gbsn}

\title{The radio/X-ray burst from SGR 1935+2154: radiation mechanisms and the possible QPOs}
\author[0000-0002-2662-6912]{Jie-Shuang Wang(王界双)}
\affiliation{Tsung-Dao Lee Institute, Shanghai Jiao Tong University, Shanghai 200240, China; jiesh.wang@gmail.com}


\begin{abstract}
Recently, a fast radio burst (FRB)-like event is found to be associated with a Galactic magnetar, SGR 1935+2154, accompanied by an X-ray burst. 
We find this radio burst challenges the typical emission mechanisms involving magnetars, which includes coherent curvature radiation from charged bunches, fast magnetosonic (FMS) wave, synchrotron maser from shocks, and the pulsar-like mechanism for low-twist magnetars. 
More specifically, we find that (1) the X-rays are most-likely to be produced inside the magnetosphere. 
(2) For the coherent curvature radiation from the decay of Alfv{\'e}n wave, it will generally predict a duration ($\lesssim0.1$ ms) smaller than observations, because of the strong twists of magnetic field lines and the internal damping of Alfv{\'e}n waves. 
(3) The FMS wave model predicts a very low emission frequency $\nu_{\rm p}\sim0.03$ MHz $\ll$ GHz, unless it is produced inside the magnetosphere. But the absorption effect of the magnetospheric FMS wave model remains to be studied. 
(4) The synchrotron maser model is challenged, because observations show that the peaks in both X-ray and radio light curves are with the same temporal separation $\Delta t_{\rm FRB}=\Delta t_\gamma\approx0.03$ s, while it would predict $\Delta t_{\rm FRB}\ll\Delta t_\gamma$. 
(5) It seems to be difficult to directly apply the low-twist pulsar-like mechanism to flaring magnetars, as magnetar activity can significantly deform the magnetosphere. 
(6) We suggested four possibilities to study the general properties of FRBs for future observations, especially the possibility of identifying quasi-periodic oscillations with period $\sim1-10$ ms  in double/multiple-peaked FRBs.
\end{abstract}

\section{Introduction}

Fast radio bursts (FRBs) are powerful radio bursts with millisecond durations. To date, more than one hundred bursts have been detected, and twenty one of them produce repeating bursts \citep[based on the catalog of][accessed in May]{Petroff2016}. 
Many models have been proposed to explain the high brightness temperate of FRBs, which usually involve magnetized compact stars as progenitors, especially neutron stars (NS) \citep[see][for a rencent review]{Platts2018}. 
In particular, it is widely suggested that young flaring magnetars \citep[\cite{Popov2013,Lyubarsky2014,Murase2016,Waxman2017,Beloborodov2017,Lu2018,Yang2018,Margalit2018,Katz2018,Beloborodov2019,Metzger2019,WangFY2020,Margalit2020,Wang2020,Kumar2020}, see also][for models associated with soft gamma-ray repeaters]{Margalit2020,Lyutikov2020,Lu2020,Yu2020,Katz2020,Yang2020,Yuan2020}, and old NSs under interaction with their companion \citep[\cite{Wang2016,Wang2018,Dai2016,Zhang2016,Zhang2020}, see also][for models associated with soft gamma-ray repeaters]{Dai2020,Geng2020} can be FRB progenitors. 

Very recently, CHIME detected a bright radio burst from the direction of a Galactic magnetar, SGR 1935+2154 \citep{Scholz2020} at $400 - 800$ MHz band with a dispersion measure DM $=332.7~{\rm pc~cm}^{-3}$. 
This powerful burst is also detected by STARE2 in $1281-1468$ MHz band with a fluence of $1.5$ MJy ms \citep{Bochenek2020atel}, which is comparable to the energy of FRBs, taking a distance of $9.1$ kpc \citep{Pavlovic2013}.
Simultaneously, the hard X-ray counterpart of this burst is reported by many high-energy observatories, for example, INTEGRAL \citep{INTEGRAL2020}, Konus-Wind \citep{Konus2020}, AGILE \citep{AGILE2020} and Insight-HXMT \citep{HXMT2020}. 
Subsequently, FAST detected anther radio pulse from this magnetar with the same DM but a fluence of 60 mJy ms \citep{FAST2020}. 
Although its fluence is much weaker than the previous one, it confirms that this magnetar is indeed in a radio active phase. 
This also supports the magnetar scenario for FRBs. 

In this paper, I propose to study the FRB mechanisms from the recent joint detection of the magnetar burst from SGR 1935+2154 by radio and X-ray observatories, and discuss the implication from the magnetar association. 
In the next section, we briefly review the observation for this event and discuss the X-ray burst model. 
In Section 3, we study the typical FRB mechanisms invoking magnetars, including coherent curvature radiation from charged bunches, fast magnetosonic (FMS) wave model, synchrotron maser from shocks, and the pulsar-like mechanism for low twist magnetars, and compare them with multi-band observations.
The predicted radio emission site by these models varies from locations close to the magnetar surface to well outside the magnetosphere. 
We discuss the possible ways to study the general properties of radio bursts for future observations in Section 4.
The summary is in Section 5.

\section{Observations and X-ray emission models}

SGR 1935+2154 is a magnetar with a spin period $P=3.245$ s \citep{Israel2016}, corresponding to a light cylinder radius $R_{\rm L}=cP/2\pi=1.55\times10^{10}$ cm. 
Its inferred surface field and spin-down luminosity are $B_{\rm s}=2.2\times10^{14}$ G and $1.7\times10^{34}$ erg/s, respectively \citep{Olausen2014}. 
The location of this magnetar by the Swift/X-ray Telescope \citep{Cummmings2014} is found to be very close to the supernova remnant G57.2+0.8 \citep{Gaensler2014}.
The latter is found to be at a distance $6.6$ kpc \citep{Zhou2020}. 
Throughout the paper, we will adopt these observational values into our calculations. 
Note although there are some other studies suggesting different source distances, e.g. $9.1$ kpc \citep{Pavlovic2013,Zhong2020} and $12.5$ kpc \citep{Kothes2018}, it does not affect our main results. 

\subsection{Multi-band observations}

For this burst detected in both X-ray and radio band, the fluence and peak flux in the $20 - 500$ keV energy band are found to be $(9.7 \pm 1.1)\times10^{-7}$ erg cm$^{-2}$ and $(7.5\pm1.0)\times10^{-6}$ erg cm$^{-2}$ s$^{-1}$, respectively \citep{Konus2020}. 
This provides the burst energy and luminosity to be $E_\gamma=(5.1\pm0.6) \times10^{39}$ erg and $L_\gamma=(3.9\pm1.0)\times 10^{40}$ erg s$^{-1}$. 
While the isotropic-equivalent radio luminosity is found to be $\sim 10^{38}$ erg/s \citep{Bochenek2020atel}, accounting for $\sim5\times10^{-3}$ of the hard X-ray burst peak luminosity. 
Both luminosities are much higher than the spin-down luminosity.

The X-ray light curve of this event shows clearly two significant pulses with a total duration of $\tau_{\gamma}\sim 0.5$ s, and the separation 
between peaks is $\Delta t_\gamma\approx30$ ms 
\citep{Konus2020,HXMT2020}. 
Interestingly, the radio light curve from CHIME also shows two peaks, 
separated by $\Delta t_{\rm FRB}=29$ ms, and each has an observed duration $\tau_{\rm FRB, obs}\sim0.5$ ms \citep{Scholz2020}. 
Further information from \cite{INTEGRAL2020} shows that there is another peak in X-rays following the two radio-associated peaks, and separated by $\sim30$ ms. 
Actually, such an additional peak can also be found in the HXMT data \citep[e.g. see Figure 2b in][]{HXMT2020}. 

We noticed that waiting times in both the X-ray and radio light curves are consistent with each other, i.e., $\Delta t_\gamma\approx\Delta t_{\rm FRB}$. 
This indicates that the $0.03$ s interval is intrinsically associated with the magnetar activity. 
Recall that observations show there are quasi-periodic oscillations (QPOs) at $\sim30$ Hz in the giant flare tails of SGR 1806$-$20 \citep{Israel2005} and SGR 1900$+$14 \citep{Strohmayer2005}. 
One might also consider that this $30$ ms interval in both X-ray and radio light curves corresponds to QPOs. 
Indeed, weak evidence of such QPOs is found in the HXMT data (Ge et al. 2020 in preparation). 
QPOs are explained as the internal oscillation modes of the magnetar, such as the crustal toroidal mode \citep[e.g.][]{Duncan1998,Piro2005}, or the magnetoelastic modes due to the crust-core coupling \citep[e.g.][]{Hoven2011,Hoven2012,Gabler2014}.

\subsection{Constraints on X-ray emission sites}\label{Xray}
Bursts from magnetars are believed to be caused by the magnetic activities, which may be initially trigger by the instabilities inside the magnetar or in the magnetosphere \citep[see][for a recent review]{Kaspi2017}. 
But it must be close to magnetar surface, as the maximum available energy dissipation rate associated with magnetic field is $L_{\rm m}=B^2(R)R^2c/2\propto R^{-4}$, where we scale the twisted magnetic field as the local dipole field $B(R)=B_{\rm s}R_{\rm NS}^3R^{-3}$ with $R_{\rm NS}$ being the NS radius. 
The maximum radius can be obtained by solving $L_{\rm m}(R_{\rm max})\approx L_\gamma$, which results in 
\beq
R_{\rm max}\approx 2.9\times 10^8L_{\gamma, 41}^{-1/4}R_{\rm NS}^{3/2}~{\rm cm},
\eeq
where $L_\gamma=L_{\gamma, 41}10^{41}$ erg/s, and $R_{\rm NS}=10^6R_{\rm NS, 6}$ cm. 
The release of magnetic energy leads to the formation of hot photon-pair plasma (a `fireball') \citep[e.g.][]{Katz1982,Katz1996,Thompson1995}, and it becomes optically thin for Thomson scattering at a radius of \citep{Meszaros2000},
\beq
R_{\rm \gamma, ph}=8.9\times10^6L_{\gamma, 41}^{1/4}R_{0,6}^{3/4}{\mathcal{Y}_2}^{1/4}~{\rm cm},\label{Rph}
\eeq
where $R_{0,6}$ is the initial size of the fireball. 
$\mathcal{Y}_2=\mathcal{Y}/10^2\sim1-10$ is the multiplicity \citep{Beloborodov2013}. 
The magnetic field at the photospheric radius is $3\times10^{11}$ G, indicating it is also optically thin for resonant scattering \citep[the resonant scattering is optically thick for $\gtrsim10^{12}$ G, see more details in][]{Beloborodov2013}. 
Thus, for this observed event, the X-rays should be emitted between $R_{\rm \gamma, ph}$ and $R_{\rm max}$.
Indeed, observations usually indicate the radiating hotspot is very compact with a size around $r_{\gamma,6}=r_{\gamma}/(10^{6}~{\rm cm})$, and the resultant luminosity is 
\beq
L_\gamma\approx \pi r_\gamma^2\sigma T^4=1.8\times10^{40}r_{\gamma,6}^2T_8^4,
\eeq
where $T_8=T/(10^8$ K) corresponds to 8.6 keV. 
Note sometimes the high-energy spectrum is different from a pure blackbody (BB) spectrum, but may be described as two BBs or Comptonization models \citep[exponential-cutoff power-laws; e.g.][]{Lin2011,Horst2012,Kaspi2017}.
This is also consistent with the observation for this event by Konus-Wind \citep{Konus2020}. 

In the following calculations, we mainly adopt that the X-rays are generated inside the magnetosphere, as $R_{\rm max}<R_{\rm L}$. 
The dynamical time of the radiation region is  much shorter than the observed duration, $R_{\rm max}/c\approx0.01$ s $\ll \tau_{\gamma}\sim0.5$ s, which indicates that the burst duration is determined by the magnetic activity timescale. 
Note this magnetospheric origins would also be supported by the possible QPOs. 
As the oscillating crust will lead to only relatively weakly periodic distortions of its attached magnetosphere \citep[e.g.][]{Gabler2014}, the existence of QPOs in light curves of giant flare tails then favours that its emission zone is compact and close to the magnetar surface \citep[e.g.][]{Timokhin2008,Thompson2017}. 
We also present a discussion about the X-rays produced outside the magnetosphere in Sec \ref{SecSM}.

\section{Radio burst models}

It is widely suggested that FRBs can be caused by magnetar activities. 
Recently, \cite{Margalit2020} studied different FRB models, mainly focused on the radiation efficiency between radios and X-rays. 
They also studied the details of synchrotron maser model with different shock properties, and favoured the case that the shock is formed through colliding with baryonic shells to explain the radiation efficiency and the X-ray counterpart. 
Here we focused on the physical conditions of different FRB mechanisms and their emission sites.

Although the radiation mechanism remains unclear, some of the mechanisms have well-defined predictions, and thus can be tested by observations. 
I study four typical mechanisms proposed for FRBs, and compare them recent observations. 
These mechanisms include coherent curvature radiation from charged bunches, fast magnetosonic (FMS) wave model, synchrotron maser from shocks, and the pulsar-like mechanism for low-twist magnetars. The emission site in these models can be close to the magnetar surface, near the light cylinder, or far away from the light cylinder.

\subsection{Coherent curvature radiation from charged bunches}\label{SecCCR}
It is widely assumed that FRBs can be generated by coherently moving electron bunches via curvature radiation \citep{Kumar2017,Lu2018,Yang2018,Kumar2020,Lu2020,Katz2020,Yang2020}, although the bunch formation and maintenance mechanisms remain unclear \citep[e.g., see][and references therein]{Melrose2017}.
Recently, \cite{Kumar2020} and \cite{Lu2020} suggested a more physical coherent curvature radiation model. 
In this model, the magnetar crustal disturbance launches a large-amplitude Alfv{\'e}n wave in the polar cap region.
The bunches are formed due to the two-stream instability of the counter-streaming pairs, and get accelerated when the Alfv{\'e}n wave becomes charge starved. 
The emission site of this model is found to be close to the surface of the neutron star $R_{\rm FRB, CC}\lesssim100R_{\rm NS}=10^8R_{\rm NS, 6}$ cm \citep{Kumar2020}. 
We consider several different cases to further constrain the emission site. 

The radiating charged bunches move at relativistic speeds, which would induce magnetic fields. 
The induced field must be small enough comparing to the local field ($B=B_{\rm s}R_{\rm NS}^3R_{\rm FRB, CC}^{-3}$) to avoid the changing of radiation direction during emission. 
This furthers requires \citep{Kumar2017,Wang2020}
\beq
R_{\rm FRB, CC}\lesssim 4\times10^7R_{\rm NS, 6}L_{\rm FRB, 38}^{1/6}~{\rm cm},\label{Rcc}
\eeq
where $L_{\rm FRB}=L_{\rm FRB, 38}10^{38}$ erg/s is the isotropic equivalent luminosity, and the magnetar surface field $B_{\rm s}=2.2\times10^{14}$ G, and radius $R_{\rm NS}=R_{\rm NS, 6}10^6$ cm are used. 
The Lorentz factor of the bunch is 
\beq
\Gamma_{\rm CC}=110 R_{\rm FRB, CC, 7}^{1/3}\nu_{\rm p, 9}^{1/3},\label{GamaCC}
\eeq 
where $\nu_{\rm p, 9}$ is the peak frequency in unit of GHz, and the curvature radius of the magnetic field line is scaled as the emission site radius $R_{\rm FRB, CC}=10^7R_{\rm FRB, CC, 7}$ cm. 
The Alfv{\'e}n wave propagation time in the radiation region is approximately $R_{\rm FRB, CC}/c\lesssim 1.3R_{\rm NS, 6}L_{\rm FRB, 38}^{1/6}~{\rm ms}$. 

However, in this model the magnetospheric Alfv{\'e}n wave is initially transmitted from crustal modes, and a high-frequency Alfv{\'e}n waves is generally required to generated the radio burst \citep{Kumar2020}. 
The wavelength is assumed to be the polar cap radius, which is $r_{\rm PC}\approx4\times 10^3$ cm for this magnetar, corresponding to a frequency $\nu_{\rm QPO}\sim10^6$ Hz. 
The twist amplitude at the magnetar surface can be calculated as 
\beq
L_\gamma\approx\delta B^2r_{\rm PC}^2c,\label{twsit}
\eeq
which gives $\delta B(R_{\rm NS,6})/B(R_{\rm NS,6})\approx10^{-3}L_{\gamma, 41}^{1/2}$ on the surface for this event. 
The twist amplitude will grow with increasing altitudes following $\delta B(R_{\rm FRB, CC,7})/B(R_{\rm FRB, CC,7})\approx0.04L_{\gamma, 41}^{1/2}R_{\rm FRB, CC, 7}^{3/2}$ \citep{Kumar2020,Yuan2020}. 
This means that the magnetic field lines are significantly deformed, leading to considerable change of the radiation direction. 
Therefore, the observed radio emission is mainly contributed by the bunches, which keeps in the line-of-sight with, i.e. $\delta B/B\lesssim1/\Gamma_{\rm CC}$. 
Combining with Eq. (\ref{GamaCC}), it further provides
\beq
R_{\rm FRB, CC}\lesssim 4\times10^6R_{\rm NS, 6}^{9/11}\nu_{\rm p, 9}^{3/22}~{\rm cm}.
\eeq
Consequently, it will lead to a duration of $0.1 R_{\rm NS, 6}^{9/11}\nu_{\rm p, 9}^{3/22}$ ms approximately. 

Additionally, such high frequency oscillation modes, if not in gaps of the Alfv{\'e}n continuum, are expected to be quickly damped due to the crust-core coupling within a time of $\sim 2\pi/\nu_{\rm QPO}=6\times10^{-6}$ s for $\nu_{\rm QPO}\sim10^6$ Hz \citep[see more details in][]{Hoven2012}. 
The plastic damping will also be significant \citep{Li2015}. 
Considering both the high twist effect and potential damping mechanisms, the resultant duration will be much smaller than the observed duration. 
Moreover, it will be very difficult to detect the low frequency ($\sim30$ ms period) QPO in this model, as it is of a wavelength of $\sim10^9$ cm, much larger than the polar cap and $R_{\rm FRB, CC}$. 

\subsection{Fast magnetosonic wave model}\label{SecFMS}
It is found that the relativistic magnetic reconnecting current sheet near the magnetosphere can generate magnetosonic waves by merging of magnetic islands. 
The fast magnetosonic (FMS) waves could successfully escape as radio waves at larger distances. 
This is suggested to be responsible for the observed radio emissions in Crab-like pulsars \citep{Lyubarsky2019,Philippov2019}. 
For FRBs converted from such FMS waves, the emission site is suggested to be close to the light cylinder \citep{Lyubarsky2020}, 
\beq
R_{\rm FRB, FMS}\sim R_{\rm L}. \label{Rfms}
\eeq
And the burst duration can be estimated as \citep[Eq. 25 in][]{Lyubarsky2020}, 
\beq 
\tau_{\rm FRB, FMS} \approx36\mathcal{Y}_2^{2/3}R_{\rm NS,6}^{1/3}L_{\gamma, 41}^{-1/2}R_{\rm esc, 15}b~{\rm ms}, \label{tfms} \eeq
where $R_{\rm esc}=R_{\rm esc, 15}10^{15}$ cm denotes the distance where FMS waves escapes freely as electromagnetic waves, and $b\lesssim1$ is the parameter relates to the mean magnetic field fraction to the total field and the magnetar wind Lorentz factor. 

Observation shows that the duration of each radio pulse is $\tau_{\rm FRB, obs}\approx0.6$ ms, which provides an escaping distance of $R_{\rm esc}\approx 10^{13}L_{\gamma, 41}^{1/2}\mathcal{Y}_2^{-2/3}R_{\rm NS,6}^{-1/3}b^{-1}$ cm for the FMS wave model by Eq. (\ref{tfms}). 
The cyclotron absorption radius in this case is found to be at $R_{\rm cy}\approx5\times10^{12}L_{\gamma, 41}^{1/2}\nu_{\rm p, GHz}^{-1}$ cm \citep[Eq. 31 in][]{Lyubarsky2020}, where $\nu_{\rm p, GHz}$ is the frequency in unit of GHz. 
This means that the wave will suffer from a mild cyclotron absorption. 

The peak frequency of the radio burst in this model is 
\beq
\nu_{\rm p, GHz}=3\times10^{-5}\eta R_{\rm NS,6}^{3/4}L_{\gamma, 41}^{5/8},
\eeq
where $\eta\sim0.1-10$ relates to details of the magnetic reconnection \citep[Eq. 12 in][]{Lyubarsky2020}. 
While this is much lower than the observed GHz frequency. 
Therefore, FMS wave model near the light cylinder is also challenged for this event, unless some extreme conditions (e.g. $\eta\sim10^5$) happens. 

However, we noticed that for this event, the radio burst might be explained if the FMS wave is generated inside the magnetosphere, rather than near the light cylinder. 
Indeed, force-free electrodynamic simulations show that magnetic reconnection could happen inside the magnetosphere, and open the closed magnetic field lines \citep{Parfrey2013,Yuan2020}. 
The advantage of the magnetospheric FMS wave model is that at a radius of $R_{\rm FMS}=R_{\rm FMS,8}10^8$ cm, the radio burst frequency will be 
\beq
\nu_{\rm p}\sim 0.2 \Gamma_p L_{\gamma, 41}^{3/4}R_{\rm FMS,8}^{-3/2}~{\rm GHz},
\eeq
where mild relativistic plasma Lorentz factor (namely $\Gamma_p\sim$ a few) is expected in the initial magnetar pulse when inside the magnetosphere. 
And the duration of the burst is $\sim R_{\rm FMS}/c=3$ ms. 
Note the FMS wave can also be produced by non-linear interactions between two Alfv{\'e}n waves \citep[e.g.][]{Li2019}. 
Such FMS wave is expected to be absorbed in the dipole magnetosphere \citep{Lyubarsky2020}. 
But whether it can escape from the highly deformed magnetosphere, where some closed field lines are opened, remains to be studied. 
We plan to study the details of this model in a future work.

\subsection{Synchrotron maser model}\label{SecSM}
Young magnetars could produce magnetar flares frequently. 
Blast waves can be launched during such flares, and their interactions with the circum-burst median can lead to the formation of magnetised shocks. 
Highly linearly polarised synchrotron maser emission can be produced by the shocks, and posse a small fraction $\sim10^{-3}$ of the total injection energy \citep{Plotnikov2019}. 
Such a process is then suggested to be the mechanism for FRBs  \citep{Lyubarsky2014,Waxman2017,Margalit2018,Beloborodov2019,Metzger2019,Margalit2020}. 
And the decelerating relativistic blast waves can well reproduce the downward frequency drifting feature observed in repeating FRBs \citep{Metzger2019}.

For FRBs produced by the synchrotron maser in relativistic shocks, the emission site is generally outside of the magnetosphere, and can be written as 
\beq
R_{\rm FRB, SM}=2\Gamma_{\rm SM}^2c\tau_{\rm FRB}=6\times10^{11}\Gamma_{\rm SM, 2}^2\tau_{\rm FRB, -3}~{\rm cm},\label{Rsm}
\eeq 
where $\Gamma_{\rm SM}=10^2\Gamma_{\rm SM, 2}$ is the bulk Lorentz factor of the shock \citep[e.g.][]{Waxman2017,Beloborodov2017,Margalit2018,Beloborodov2019}, $\tau_{\rm FRB}=10^{-3}\tau_{\rm FRB, -3}$ s is the burst duration. 
For the shock developed outside the magnetosphere $R>R_{\rm L}$, a bulk Lorentz factor of $\Gamma_{\rm SM}^2>R_{\rm L}/(2c\tau_{\rm FRB})\approx250\tau_{\rm FRB, -3}^{-1}$ is required to justify the observed FRB duration. 
We note in this case it is unlikely to produce double/multiple peaks, especially the QPO behaviour, in the X-ray and radio light curve by one blast wave.
Therefore, multiple blast waves are required. 
Considering the magnetar generates several X-ray burst peaks with a time separation of $\Delta t$, the associated blast waves would form shocks and thus lead to the peak separation in the radio band to be $\Delta t_{\rm FRB}=0.5\Gamma_{\rm SM}^{-2}\Delta t_\gamma\ll\Delta t_\gamma$, much smaller than that in X-rays.
This is incompatible with the observed delay between the radio peaks, which is $\Delta t_{\rm FRB}\approx\Delta t_\gamma\approx30$ ms.

The above calculations are on the basis that X-rays are produced inside the magnetosphere (see Section \ref{Xray}). 
We noted that it is also suggested that the relativistic electrons in the shock can also produce X-ray emissions \citep{Metzger2019,Margalit2020,Yu2020,Yuan2020}. 
However, the predicted duration of the X-ray emission from the shock is $\mathcal{O}(\tau_{\rm FRB})=\mathcal{O}(1~{\rm ms})$, which is inconsistent with the observed duration of the X-ray pulse $\sim50$ ms $\gg1$ ms \citep{Konus2020,HXMT2020}. 
In particular, within the framework of shock model it will be difficult to explain the observed slow rise of the X-ray pulse, the QPO behaviours, and the null detection of radio signals associated with the third X-ray pulse. 

Moreover, we find that the shock model cannot explain the observed synchrotron peak energies from the two X-ray pulses in the light curve either. 
As the first shock will sweep the circum-burst median, the second one is formed when colliding with the decelerated tail of the first shock. 
As such the two radio pulses separated by $30$ ms would lead to that two shocks forming at different distance. 
\cite{Lu2020} found the Lorenz factors of two shocks (denoted by subscript $1,~2$) differ by $\Gamma_{\rm SM, 2}/\Gamma_{\rm SM, 1}\sim5$, while the external densities differ by $n_2/n_1\sim0.04$. 
The synchrotron energy scales as $\nu_{\rm syn}\propto n^{1/2}\Gamma_{\rm SM}^4 $ \citep{Metzger2019}. 
The resultant synchrotron peak energies between two pulses is $\nu_{\rm syn, 2}/ \nu_{\rm syn, 1} \sim 100$.
This means that if the first X-ray pulse is peaked at $100$ keV, the second pulse will be peaked at $\sim1$ keV, which is clearly inconsistent with observations \citep{Konus2020,HXMT2020}. 
To conclude, it is unlikely that the incoherent synchrotron radiation of the shock will significantly contribute to the observed X-rays.  

\subsection{Pulsar-like mechanism with low magnetospheric twists}
Recently, it is also suggested that repeating FRBs can be produced by pulsar-like mechanisms, although the detailed radiation mechanism is not specified \citep{Wadiasingh2019}. 
This model requires a pre-existing low twist of the magnetosphere, and then disturbed by the magnetar burst activities. 
The shift of foot points of magnetic field lines induces an acceleration gap, acting like the polar cap in pulsars. 
And thus pulsar-like radio emissions are assumed to be produced. 
Within such a framework, \cite{Wadiasingh2020} further predicted a death line for FRBs, which says that only magnetars with period $P\gtrsim0.2$ s and magnetic field $B\gtrsim 6\times 10^{13}P^{-1}$ G can generate FRBs. 
SGR 1935+2154 well meets this requirement, with $P=3.245$ s and $B=2.2\times10^{14}$ G. 
In this model, the radio emission continues until the pair plasma screens the acceleration gap with a timescale $\sim R/c= 3.3 R_8$ ms for closed fields with maximum altitude $R=R_810^8$ cm \citep{Wadiasingh2019}. 

However, the plasma would screen the acceleration field after one event, and the time required to clear this plasma is much longer than the burst duration \citep{Wadiasingh2019}. 
Therefore, it will be difficult to produce double and multi peaked burst. 
Moreover, the crustal oscillation induced shift of foot points can strongly disturb the magnetic field lines. 
Like the discussion in Section \ref{SecCCR}, although the initial twist angle is quite small $\delta B/B\sim8\times10^{-4}L_{\gamma, 41}^{1/2}r_{\rm FP,4}$ (Eq. \ref{twsit}), it grows with increasing altitudes as $\delta B/B\sim0.5r_{\rm FP,4}R_8^{3/2}$, where the polar cap size is replaced by the size of regions with disturbed field foot points $r_{\rm FP}=10^4r_{\rm FP,4}$ cm. 
Consequently, if the radiation is connected to the cascaded plasma along magnetic field lines \citep[as assumed by][]{Wadiasingh2019}, the radiation direction would change significantly during the burst, and thus leads to a much shorter duration. 
Furthermore, the deformed magnetosphere is no longer suitable for the pulsar-like mechanisms, as the twist grows much higher. 

\section{Discussion on constraining radio burst properties with future observations}

Although the radio emission mechanism remains unknown, we here discuss some possible general ways to study the properties of radio bursts for future observations.  
The above models will generally predict a time difference between the X-ray and radio peaks to be $R/(2\Gamma_{\rm SM}^2c)\sim\tau_{\rm FRB}\approx1$ ms for the synchrotron maser model or $<R_{\rm max}/c\approx10$ ms for other models. 
Observations from CHIME show that the two radio peaks arrive at dispersion-corrected times 14:34:24.4285 and 14:34:24.4575 on Apr. 28, 2020 \citep{Scholz2020}. While observation shows that the two X-ray peaks come at 14:34:24.428$\pm0.002$ and 14:34:24.460$\pm0.002$ \citep{HXMT2020}, which indicate an intrinsic time delay of $\delta t_{\rm d}\leq5$ ms. Similar results are also found by Konus-Wind \citep{Konus2020}. 
This result is consistent with all models. 
However, if future detections show that the X-ray and radio burst could be separated with a much longer timescale (e.g. $\gg10$ ms), these models would be challenged. 

While the association between FRB-like burst and magnetar burst is firmly established, it is still unclear whether all X-ray bursts are associated with radio bursts. 
X-ray observation shows that the energy of this burst is quite normal comparing with other bursts without association of radio bursts, but its spectrum appears to be much harder \citep{INTEGRAL2020,Konus2020}. 
We here consider one possibility under the dual BB framework. 
The hard X-rays in this case are explained by a hotter BB component but with a smaller radiation area than that of soft X-rays \citep{Lin2011,Horst2012}, which is also consistent with the observation for this event by Konus-Wind \citep{Konus2020}. 
Thus, the opening angle of hard X-ray radiation will also be smaller than that of soft X-rays, and if it offsets from line-of-sight, one might only detect the soft X-ray component. 
If future detections indeed support the connection between hard X-rays and radio bursts, this would strongly suggest that their radiation cones point very close to each other.
Note the opening angles of X-ray and radio emissions can also be very different. 
One way to estimate it is to assume that the observed event rate density correlates to opening angle, namely, $\Delta \Omega_\gamma/\Delta \Omega_{\rm FRB}\sim \dot{N}_\gamma/\dot{N}_{\rm FRB}$, which can be calculated if more data is available. This is also suggested by a recent study \citep{Lin2020uplimit}.

We note the flaring activity of magnetars could also have other counterparts, if there are outflows associated with the activity. 
For repeating FRBs, intermittent outflows can be injected to the circum-burst median. 
Note the shock produced by such an outflow does not necessarily generate significant coherent radio emission, e.g. the shock in pulsar wind nebula. 
If the waiting time between two outflows is much smaller than the cooling time of the electrons accelerated by the outflow, an FRB nebula will form, like the pulsar wind nebula. 
Otherwise, it forms an FRB afterglow. 
Such an FRB nebula can well explain the persistent radio counterpart of FRB 121102 \citep[e.g.][]{Waxman2017,Beloborodov2017,Margalit2018,Wang2020}. 
More specifically, \cite{Wang2020} provided a tight constraint on energetic parameters of the central engine from the radio nebula for FRB 121102. 
Additionally, \cite{Wang2020} also studied the multi-band light curves and spectra of the afterglow. 
Although our calculations show that for this event the radio afterglow is too faint to be observed, one might expect to detect such afterglows for much stronger bursts in the future. 

There is weak evidence showing that the X-ray and radio light curves are modulated by low-frequency QPOs with period $\sim\mathcal{O}(10$ ms), which can be related to fundamental oscillation modes of the magnetar. 
Similar to those in giant flare tails, the existence of QPOs would also prefer that these emissions are most likely emerging from inside the magnetosphere. 
We noted that if FRBs are generated close to the magnetar surface, there is also a possibility to detect high-frequency QPOs with period $0.5 -2$ ms.
Such high frequency QPOs (e.g. 625 Hz and 1840 Hz) have been detected in giant flare tails, which can be caused by the overtones of oscillation modes \citep[e.g.][]{Watts2007}. 
Interestingly, this would show as double/multiple peaks in FRBs. 
Therefore, it will be interesting to test if some of the double/multi-peaked FRB are caused by the QPO effect.

\section{Summary}

In this paper, we constrain the radio emission mechanisms and the emission sites for the magnetar burst from SGR 1935+2154, which is recently detected in both X-ray and radio bands. 
More specifically, we compare observations with the predictions of the relatively well-studied FRB radiation mechanisms, including coherent curvature radiation, escaped FMS wave, synchrotron maser, and pulsar-like mechanisms.
The main results are summarised as follows.
\begin{enumerate}
	\item The X-ray emission radius is constrained to be between $R_{\rm \gamma, ph}=8.9\times10^6$ cm and $R_{\rm max}\approx 2.9\times 10^8$ cm. 
	This leads to a dynamical timescale, much shorter than the observed duration $R_{\rm max}/c\ll0.5$ s. 
	As such, the X-ray burst duration should be directly determined by the magnetar activity timescale. 
	\item For the coherent curvature radiation induced by the decay of  Alfv{\'e}n waves, we find that the radiation region should be very close to the magnetar surface, namely $R_{\rm FRB, CC}\lesssim 4\times10^6R_{\rm NS, 6}^{9/11}\nu_{\rm p, 9}^{3/22}~{\rm cm}$, with a duration ($R_{\rm FRB, CC}/c\lesssim0.1$ ms) smaller than the observed duration. 
	Besides, such a high-frequency ($10^6$ Hz) crustal mode is most likely to be damped within a timescale of $\sim6\times10^{-6}$ s due to the crust-core coupling, much shorter than the burst duration.  
	Moreover, in this case the burst most-likely will not be affected by low-frequency waves (e.g. $30$ ms QPO), which is inconsistent with observations. 
	Therefore, mechanisms for bunch formation and acceleration remain to be studied to make the coherent curvature radiation viable. 
	\item For the FMS wave model, we find that the predicted emission frequency ($\nu_{\rm p}=3\times10^{-5}\eta $ GHz) is found to be much smaller than the observed GHz emission, unless the magnetic reconnection take places under some extreme conditions with $\eta\sim10^5$ or inside the magnetosphere. 
	For the latter case, however, the absorption effect remains to be studied.
	Besides, the escaping wave will suffer from a mild absorption, as the cyclotron absorption distance is smaller than escaping distance $R_{\rm cy}\approx 5\times10^{12}$ cm $\lesssim R_{\rm esc}\approx10^{13}$ cm. 
	\item For the synchrotron maser developed outside the light cylinder, it would generally predict $\Delta t_{\rm FRB}/\Delta t_\gamma=0.5\Gamma_{\rm SM}^{-2}\lesssim10^{-3}\tau_{\rm FRB, -3}$. 
	While observations show that the two peaks in both the X-ray and radio light curves are of the same temporal separation $\Delta t_\gamma\approx\Delta t_{\rm FRB}\approx 30$ ms. 
	Therefore, synchrotron maser is disfavoured, unless that the X-rays are also contributed by the electrons of the shock. 
	However, the shock model for X-rays will be difficult to explain the observed duration (especially the slow rise), the QPO behaviours of the X-ray pulse and the null detection of radio signals associated with the third X-ray pulse. 
	Furthermore, the X-rays from the shock would have a significant spectral evolution between pulses, which are inconsistent with observations. 
	\item For the pulsar-like mechanisms, most-likely it is unable to produce double/multiple-peaked bursts in flaring magnetars. 
	Furthermore, we find that it is actually difficult to directly apply pulsar-like mechanisms \citep[described in][]{Wadiasingh2019} to magnetar bursts, because that the pre-existing low twist magnetosphere can be significantly deformed during the burst activity in flaring magnetars, which is unlike the polar caps in pulsars. 
	But a solid conclusion can not be drawn for now, as the radiation mechanisms for both FRBs and pulsars remain enigmatic. 
	\item We also propose four ways to study the radio burst properties for future observations, including the time delay between X-ray and radio bursts, the opening angle of the radio burst, the FRB afterglow or nebula, and the possible QPOs in radio bursts. 
	For the latter case, it will be interesting to test the possibility of identifying QPOs, including both high-frequency with period $\sim\mathcal{O}(1$ ms)  and low-frequency with period $\sim\mathcal{O}(10$ ms) QPOs in double/multiple-peaked FRBs. 
\end{enumerate}

To conclude, although these models are suggested to be associated with X-ray bursts, as they are induced by flaring activities, our constraints indicate that further researches are required to explain the details of both the X-ray and radio observations, especially the temporal (e.g. QPOs) behaviours of the bursts and the physical requirements for radio emission mechanisms.

\acknowledgements
We thank the referee for useful comments, and Yuanpei Yang, Xinyu Li, Yunwei Yu, Ben Margalit, Dong Lai, and Lingjun Wang for fruitful discussions. 
JSW is supported by China Postdoctoral Science Foundation (Grant 2018M642000, 2019T120335).

\bibliographystyle{aasjournal}
\bibliography{ref}

\begin{thebibliography}{}
\expandafter\ifx\csname natexlab\endcsname\relax\def\natexlab#1{#1}\fi
\providecommand{\url}[1]{\href{#1}{#1}}
\providecommand{\dodoi}[1]{doi:~\href{http://doi.org/#1}{\nolinkurl{#1}}}
\providecommand{\doeprint}[1]{\href{http://ascl.net/#1}{\nolinkurl{http://ascl.net/#1}}}
\providecommand{\doarXiv}[1]{\href{https://arxiv.org/abs/#1}{\nolinkurl{https://arxiv.org/abs/#1}}}

\bibitem[{{Beloborodov}(2013)}]{Beloborodov2013}
{Beloborodov}, A.~M. 2013, \apj, 762, 13, \dodoi{10.1088/0004-637X/762/1/13}

\bibitem[{{Beloborodov}(2017)}]{Beloborodov2017}
---. 2017, \apj, 843, L26, \dodoi{10.3847/2041-8213/aa78f3}

\bibitem[{{Beloborodov}(2019)}]{Beloborodov2019}
---. 2019, arXiv e-prints, arXiv:1908.07743.
\newblock \doarXiv{1908.07743}

\bibitem[{{Bochenek} {et~al.}(2020){Bochenek}, {Ravi}, {Belov}, {Hallinan},
  {Kocz}, {Kulkarni}, \& {McKenna}}]{Bochenek2020atel}
{Bochenek}, C.~D., {Ravi}, V., {Belov}, K.~V., {et~al.} 2020, arXiv e-prints,
  arXiv:2005.10828.
\newblock \doarXiv{2005.10828}

\bibitem[{{Cummmings} {et~al.}(2014){Cummmings}, {Barthelmy}, {Chester}, \&
  {Page}}]{Cummmings2014}
{Cummmings}, J.~R., {Barthelmy}, S.~D., {Chester}, M.~M., \& {Page}, K.~L.
  2014, The Astronomer's Telegram, 6294, 1

\bibitem[{{Dai}(2020)}]{Dai2020}
{Dai}, Z.~G. 2020, arXiv e-prints, arXiv:2005.12048.
\newblock \doarXiv{2005.12048}

\bibitem[{{Dai} {et~al.}(2016){Dai}, {Wang}, {Wu}, \& {Huang}}]{Dai2016}
{Dai}, Z.~G., {Wang}, J.~S., {Wu}, X.~F., \& {Huang}, Y.~F. 2016, \apj, 829,
  27, \dodoi{10.3847/0004-637X/829/1/27}

\bibitem[{{Duncan}(1998)}]{Duncan1998}
{Duncan}, R.~C. 1998, \apjl, 498, L45, \dodoi{10.1086/311303}

\bibitem[{{Gabler} {et~al.}(2014){Gabler}, {Cerd{\'a}-Dur{\'a}n},
  {Stergioulas}, {Font}, \& {M{\"u}ller}}]{Gabler2014}
{Gabler}, M., {Cerd{\'a}-Dur{\'a}n}, P., {Stergioulas}, N., {Font}, J.~A., \&
  {M{\"u}ller}, E. 2014, \mnras, 443, 1416, \dodoi{10.1093/mnras/stu1263}

\bibitem[{{Gaensler}(2014)}]{Gaensler2014}
{Gaensler}, B.~M. 2014, GRB Coordinates Network, 16533, 1

\bibitem[{{Geng} {et~al.}(2020){Geng}, {Li}, {Li}, {Xiong}, {Kuiper}, \&
  {Huang}}]{Geng2020}
{Geng}, J.-J., {Li}, B., {Li}, L.-B., {et~al.} 2020, arXiv e-prints,
  arXiv:2006.04601.
\newblock \doarXiv{2006.04601}

\bibitem[{{Israel} {et~al.}(2005){Israel}, {Belloni}, {Stella}, {Rephaeli},
  {Gruber}, {Casella}, {Dall'Osso}, {Rea}, {Persic}, \&
  {Rothschild}}]{Israel2005}
{Israel}, G.~L., {Belloni}, T., {Stella}, L., {et~al.} 2005, \apjl, 628, L53,
  \dodoi{10.1086/432615}

\bibitem[{{Israel} {et~al.}(2016){Israel}, {Esposito}, {Rea}, {Coti Zelati},
  {Tiengo}, {Campana}, {Mereghetti}, {Rodriguez Castillo}, {G{\"o}tz},
  {Burgay}, {Possenti}, {Zane}, {Turolla}, {Perna}, {Cannizzaro}, \&
  {Pons}}]{Israel2016}
{Israel}, G.~L., {Esposito}, P., {Rea}, N., {et~al.} 2016, \mnras, 457, 3448,
  \dodoi{10.1093/mnras/stw008}

\bibitem[{{Kaspi} \& {Beloborodov}(2017)}]{Kaspi2017}
{Kaspi}, V.~M., \& {Beloborodov}, A.~M. 2017, \araa, 55, 261,
  \dodoi{10.1146/annurev-astro-081915-023329}

\bibitem[{{Katz}(1982)}]{Katz1982}
{Katz}, J.~I. 1982, \apj, 260, 371, \dodoi{10.1086/160262}

\bibitem[{{Katz}(1996)}]{Katz1996}
---. 1996, \apj, 463, 305, \dodoi{10.1086/177242}

\bibitem[{{Katz}(2018)}]{Katz2018}
---. 2018, Progress in Particle and Nuclear Physics, 103, 1,
  \dodoi{10.1016/j.ppnp.2018.07.001}

\bibitem[{{Katz}(2020)}]{Katz2020}
---. 2020, arXiv e-prints, arXiv:2006.03468.
\newblock \doarXiv{2006.03468}

\bibitem[{{Kothes} {et~al.}(2018){Kothes}, {Sun}, {Gaensler}, \&
  {Reich}}]{Kothes2018}
{Kothes}, R., {Sun}, X., {Gaensler}, B., \& {Reich}, W. 2018, \apj, 852, 54,
  \dodoi{10.3847/1538-4357/aa9e89}

\bibitem[{{Kumar} \& {Bo{\v{s}}njak}(2020)}]{Kumar2020}
{Kumar}, P., \& {Bo{\v{s}}njak}, {\v{Z}}. 2020, \mnras, 494, 2385,
  \dodoi{10.1093/mnras/staa774}

\bibitem[{{Kumar} {et~al.}(2017){Kumar}, {Lu}, \& {Bhattacharya}}]{Kumar2017}
{Kumar}, P., {Lu}, W., \& {Bhattacharya}, M. 2017, \mnras, 468, 2726,
  \dodoi{10.1093/mnras/stx665}

\bibitem[{{Li} {et~al.}(2020){Li}, {Lin}, {Xiong}, {Ge}, {Li}, {Li}, {Lu},
  {Zhang}, {Tuo}, {Nang}, {Zhang}, {Xiao}, {Chen}, {Song}, {Xu}, {Liu}, {Jia},
  {Cao}, {Zhang}, {Qu}, {Liao}, {Zhao}, {Tan}, {Nie}, {Zhao}, {Zheng}, {Zheng},
  {Luo}, {Cai}, {Li}, {Xue}, {Bu}, {Chang}, {Chen}, {Chen}, {Chen}, {Chen},
  {Chen}, {Cui}, {Cui}, {Deng}, {Dong}, {Du}, {Fu}, {Gao}, {Gao}, {Gao}, {Gu},
  {Guan}, {Guo}, {Han}, {Huang}, {Huo}, {Jiang}, {Jiang}, {Jin}, {Jin}, {Kong},
  {Li}, {Li}, {Li}, {Li}, {Li}, {Li}, {Li}, {Liang}, {Liu}, {Liu}, {Liu},
  {Liu}, {Liu}, {Lu}, {Lu}, {Luo}, {Ma}, {Meng}, {Ou}, {Sai}, {Shang}, {Song},
  {Sun}, {Tao}, {Wang}, {Wang}, {Wang}, {Wang}, {Wang}, {Wen}, {Wu}, {Wu},
  {Wu}, {Xiao}, {Yang}, {Yang}, {Yang}, {Yang}, {Yi}, {Yin}, {You}, {Zhang},
  {Zhang}, {Zhang}, {Zhang}, {Zhang}, {Zhang}, {Zhang}, {Zhang}, {Zhang},
  {Zhang}, {Zhang}, {Zhang}, {Zhang}, {Zhang}, {Zhang}, {Zhang}, {Zhou},
  {Zhou}, {Zhu}, {Zhu}, \& {Zhuang}}]{HXMT2020}
{Li}, C.~K., {Lin}, L., {Xiong}, S.~L., {et~al.} 2020, arXiv e-prints,
  arXiv:2005.11071.
\newblock \doarXiv{2005.11071}

\bibitem[{{Li} \& {Beloborodov}(2015)}]{Li2015}
{Li}, X., \& {Beloborodov}, A.~M. 2015, \apj, 815, 25,
  \dodoi{10.1088/0004-637X/815/1/25}

\bibitem[{{Li} {et~al.}(2019){Li}, {Zrake}, \& {Beloborodov}}]{Li2019}
{Li}, X., {Zrake}, J., \& {Beloborodov}, A.~M. 2019, \apj, 881, 13,
  \dodoi{10.3847/1538-4357/ab2a03}

\bibitem[{{Lin} {et~al.}(2011){Lin}, {Kouveliotou}, {Baring}, {van der Horst},
  {Guiriec}, {Woods}, {G{\"o}{\v{g}}{\"u}{\textcommabelow s}}, {Kaneko},
  {Scargle}, {Granot}, {Preece}, {von Kienlin}, {Chaplin}, {Watts}, {Wijers},
  {Zhang}, {Bhat}, {Finger}, {Gehrels}, {Harding}, {Kaper}, {Kaspi}, {Mcenery},
  {Meegan}, {Paciesas}, {Pe'er}, {Ramirez-Ruiz}, {van der Klis}, {Wachter}, \&
  {Wilson-Hodge}}]{Lin2011}
{Lin}, L., {Kouveliotou}, C., {Baring}, M.~G., {et~al.} 2011, \apj, 739, 87,
  \dodoi{10.1088/0004-637X/739/2/87}

\bibitem[{{Lin} {et~al.}(2020){Lin}, {Zhang}, {Wang}, {Gao}, {Guan}, {Han},
  {Jiang}, {Jiang}, {Lee}, {Li}, {Men}, {Miao}, {Niu}, {Niu}, {Sun}, {Wang},
  {Wang}, {Xu}, {Xu}, {Xu}, {Yang}, {Yang}, {Yu}, {Zhang}, {Zhang}, {Zhou},
  {Zhu}, {Castro-Tirado}, {Dai}, {Ge}, {Hu}, {Li}, {Li}, {Li}, {Liang}, {Jia},
  {Querel}, {Shao}, {Wang}, {Wang}, {Wu}, {Xiong}, {Xu}, {Yang}, {Zhang},
  {Zhang}, {Zheng}, \& {Zou}}]{Lin2020uplimit}
{Lin}, L., {Zhang}, C.~F., {Wang}, P., {et~al.} 2020, arXiv e-prints,
  arXiv:2005.11479.
\newblock \doarXiv{2005.11479}

\bibitem[{{Lu} \& {Kumar}(2018)}]{Lu2018}
{Lu}, W., \& {Kumar}, P. 2018, \mnras, 477, 2470, \dodoi{10.1093/mnras/sty716}

\bibitem[{{Lu} {et~al.}(2020){Lu}, {Kumar}, \& {Zhang}}]{Lu2020}
{Lu}, W., {Kumar}, P., \& {Zhang}, B. 2020, arXiv e-prints, arXiv:2005.06736.
\newblock \doarXiv{2005.06736}

\bibitem[{{Lyubarsky}(2014)}]{Lyubarsky2014}
{Lyubarsky}, Y. 2014, \mnras, 442, L9, \dodoi{10.1093/mnrasl/slu046}

\bibitem[{{Lyubarsky}(2019)}]{Lyubarsky2019}
---. 2019, \mnras, 483, 1731, \dodoi{10.1093/mnras/sty3233}

\bibitem[{{Lyubarsky}(2020)}]{Lyubarsky2020}
---. 2020, arXiv e-prints, arXiv:2001.02007.
\newblock \doarXiv{2001.02007}

\bibitem[{{Lyutikov} \& {Popov}(2020)}]{Lyutikov2020}
{Lyutikov}, M., \& {Popov}, S. 2020, arXiv e-prints, arXiv:2005.05093.
\newblock \doarXiv{2005.05093}

\bibitem[{{Margalit} {et~al.}(2020){Margalit}, {Beniamini}, {Sridhar}, \&
  {Metzger}}]{Margalit2020}
{Margalit}, B., {Beniamini}, P., {Sridhar}, N., \& {Metzger}, B.~D. 2020, arXiv
  e-prints, arXiv:2005.05283.
\newblock \doarXiv{2005.05283}

\bibitem[{{Margalit} \& {Metzger}(2018)}]{Margalit2018}
{Margalit}, B., \& {Metzger}, B.~D. 2018, \apj, 868, L4,
  \dodoi{10.3847/2041-8213/aaedad}

\bibitem[{{Melrose}(2017)}]{Melrose2017}
{Melrose}, D.~B. 2017, Reviews of Modern Plasma Physics, 1, 5,
  \dodoi{10.1007/s41614-017-0007-0}

\bibitem[{{Mereghetti} {et~al.}(2020){Mereghetti}, {Savchenko}, {Ferrigno},
  {G{\"o}tz}, {Rigoselli}, {Tiengo}, {Bazzano}, {Bozzo}, {Coleiro},
  {Courvoisier}, {Doyle}, {Goldwurm}, {Hanlon}, {Jourdain}, {von Kienlin},
  {Lutovinov}, {Martin-Carrillo}, {Molkov}, {Natalucci}, {Onori}, {Panessa},
  {Rodi}, {Rodriguez}, {S{\'a}nchez-Fern{\'a}ndez}, {Sunyaev}, \&
  {Ubertini}}]{INTEGRAL2020}
{Mereghetti}, S., {Savchenko}, V., {Ferrigno}, C., {et~al.} 2020, arXiv
  e-prints, arXiv:2005.06335.
\newblock \doarXiv{2005.06335}

\bibitem[{{M{\'e}sz{\'a}ros} \& {Rees}(2000)}]{Meszaros2000}
{M{\'e}sz{\'a}ros}, P., \& {Rees}, M.~J. 2000, \apj, 530, 292,
  \dodoi{10.1086/308371}

\bibitem[{{Metzger} {et~al.}(2019){Metzger}, {Margalit}, \&
  {Sironi}}]{Metzger2019}
{Metzger}, B.~D., {Margalit}, B., \& {Sironi}, L. 2019, arXiv e-prints.
\newblock \doarXiv{1902.01866}

\bibitem[{{Murase} {et~al.}(2016){Murase}, {Kashiyama}, \&
  {M{\'e}sz{\'a}ros}}]{Murase2016}
{Murase}, K., {Kashiyama}, K., \& {M{\'e}sz{\'a}ros}, P. 2016, \mnras, 461,
  1498, \dodoi{10.1093/mnras/stw1328}

\bibitem[{{Olausen} \& {Kaspi}(2014)}]{Olausen2014}
{Olausen}, S.~A., \& {Kaspi}, V.~M. 2014, \apjs, 212, 6,
  \dodoi{10.1088/0067-0049/212/1/6}

\bibitem[{{Parfrey} {et~al.}(2013){Parfrey}, {Beloborodov}, \&
  {Hui}}]{Parfrey2013}
{Parfrey}, K., {Beloborodov}, A.~M., \& {Hui}, L. 2013, \apj, 774, 92,
  \dodoi{10.1088/0004-637X/774/2/92}

\bibitem[{{Pavlovi{\'c}} {et~al.}(2013){Pavlovi{\'c}}, {Uro{\v{s}}evi{\'c}},
  {Vukoti{\'c}}, {Arbutina}, \& {G{\"o}ker}}]{Pavlovic2013}
{Pavlovi{\'c}}, M.~Z., {Uro{\v{s}}evi{\'c}}, D., {Vukoti{\'c}}, B., {Arbutina},
  B., \& {G{\"o}ker}, {\"U}.~D. 2013, \apjs, 204, 4,
  \dodoi{10.1088/0067-0049/204/1/4}

\bibitem[{{Petroff} {et~al.}(2016){Petroff}, {Barr}, {Jameson}, {Keane},
  {Bailes}, {Kramer}, {Morello}, {Tabbara}, \& {van Straten}}]{Petroff2016}
{Petroff}, E., {Barr}, E.~D., {Jameson}, A., {et~al.} 2016, \pasa, 33, e045,
  \dodoi{10.1017/pasa.2016.35}

\bibitem[{{Philippov} {et~al.}(2019){Philippov}, {Uzdensky}, {Spitkovsky}, \&
  {Cerutti}}]{Philippov2019}
{Philippov}, A., {Uzdensky}, D.~A., {Spitkovsky}, A., \& {Cerutti}, B. 2019,
  \apjl, 876, L6, \dodoi{10.3847/2041-8213/ab1590}

\bibitem[{{Piro}(2005)}]{Piro2005}
{Piro}, A.~L. 2005, \apjl, 634, L153, \dodoi{10.1086/499049}

\bibitem[{{Platts} {et~al.}(2018){Platts}, {Weltman}, {Walters}, {Tendulkar},
  {Gordin}, \& {Kandhai}}]{Platts2018}
{Platts}, E., {Weltman}, A., {Walters}, A., {et~al.} 2018, arXiv e-prints,
  arXiv:1810.05836.
\newblock \doarXiv{1810.05836}

\bibitem[{{Plotnikov} \& {Sironi}(2019)}]{Plotnikov2019}
{Plotnikov}, I., \& {Sironi}, L. 2019, \mnras, \dodoi{10.1093/mnras/stz640}

\bibitem[{{Popov} \& {Postnov}(2013)}]{Popov2013}
{Popov}, S.~B., \& {Postnov}, K.~A. 2013, arXiv e-prints, arXiv:1307.4924.
\newblock \doarXiv{1307.4924}

\bibitem[{{Ridnaia} {et~al.}(2020){Ridnaia}, {Svinkin}, {Frederiks}, {Bykov},
  {Popov}, {Aptekar}, {Golenetskii}, {Lysenko}, {Tsvetkova}, {Ulanov}, \&
  {Cline}}]{Konus2020}
{Ridnaia}, A., {Svinkin}, D., {Frederiks}, D., {et~al.} 2020, arXiv e-prints,
  arXiv:2005.11178.
\newblock \doarXiv{2005.11178}

\bibitem[{{Strohmayer} \& {Watts}(2005)}]{Strohmayer2005}
{Strohmayer}, T.~E., \& {Watts}, A.~L. 2005, \apjl, 632, L111,
  \dodoi{10.1086/497911}

\bibitem[{{Tavani} {et~al.}(2020){Tavani}, {Casentini}, {Ursi}, {Verrecchia},
  {Addis}, {Antonelli}, {Argan}, {Barbiellini}, {Baroncelli}, {Bernardi},
  {Bianchi}, {Bulgarelli}, {Caraveo}, {Cardillo}, {Cattaneo}, {Chen}, {Costa},
  {Del Monte}, {Di Cocco}, {Di Persio}, {Donnarumma}, {Evangelista}, {Feroci},
  {Ferrari}, {Fioretti}, {Fuschino}, {Galli}, {Gianotti}, {Giuliani},
  {Labanti}, {Lazzarotto}, {Lipari}, {Longo}, {Lucarelli}, {Magro},
  {Marisaldi}, {Mereghetti}, {Morelli}, {Morselli}, {Naldi}, {Pacciani},
  {Parmiggiani}, {Paoletti}, {Pellizzoni}, {Perri}, {Perotti}, {Piano},
  {Picozza}, {Pilia}, {Pittori}, {Puccetti}, {Pupillo}, {Rapisarda},
  {Rappoldi}, {Rubini}, {Setti}, {Soffitta}, {Trifoglio}, {Trois},
  {Vercellone}, {Vittorini}, {Giommi}, \& {D' Amico}}]{AGILE2020}
{Tavani}, M., {Casentini}, C., {Ursi}, A., {et~al.} 2020, arXiv e-prints,
  arXiv:2005.12164.
\newblock \doarXiv{2005.12164}

\bibitem[{{The CHIME/FRB Collaboration} {et~al.}(2020){The CHIME/FRB
  Collaboration}, {:}, {Andersen}, {Band ura}, {Bhardwaj}, {Bij}, {Boyce},
  {Boyle}, {Brar}, {Cassanelli}, {Chawla}, {Chen}, {Cliche}, {Cook},
  {Cubranic}, {Curtin}, {Denman}, {Dobbs}, {Dong}, {Fandino}, {Fonseca},
  {Gaensler}, {Giri}, {Good}, {Halpern}, {Hill}, {Hinshaw}, {H{\"o}fer},
  {Josephy}, {Kania}, {Kaspi}, {Landecker}, {Leung}, {Li}, {Lin}, {Masui},
  {Mckinven}, {Mena-Parra}, {Merryfield}, {Meyers}, {Michilli}, {Milutinovic},
  {Mirhosseini}, {M{\"u}nchmeyer}, {Naidu}, {Newburgh}, {Ng}, {Patel}, {Pen},
  {Pinsonneault-Marotte}, {Pleunis}, {Quine}, {Rafiei-Ravandi}, {Rahman},
  {Ransom}, {Renard}, {Sanghavi}, {Scholz}, {Shaw}, {Shin}, {Siegel}, {Singh},
  {Smegal}, {Smith}, {Stairs}, {Tan}, {Tendulkar}, {Tretyakov}, {Vanderlinde},
  {Wang}, {Wulf}, \& {Zwaniga}}]{Scholz2020}
{The CHIME/FRB Collaboration}, {:}, {Andersen}, B.~C., {et~al.} 2020, arXiv
  e-prints, arXiv:2005.10324.
\newblock \doarXiv{2005.10324}

\bibitem[{{Thompson} \& {Duncan}(1995)}]{Thompson1995}
{Thompson}, C., \& {Duncan}, R.~C. 1995, \mnras, 275, 255,
  \dodoi{10.1093/mnras/275.2.255}

\bibitem[{{Thompson} {et~al.}(2017){Thompson}, {Yang}, \&
  {Ortiz}}]{Thompson2017}
{Thompson}, C., {Yang}, H., \& {Ortiz}, N. 2017, \apj, 841, 54,
  \dodoi{10.3847/1538-4357/aa6c30}

\bibitem[{{Timokhin} {et~al.}(2008){Timokhin}, {Eichler}, \&
  {Lyubarsky}}]{Timokhin2008}
{Timokhin}, A.~N., {Eichler}, D., \& {Lyubarsky}, Y. 2008, \apj, 680, 1398,
  \dodoi{10.1086/587925}

\bibitem[{{van der Horst} {et~al.}(2012){van der Horst}, {Kouveliotou},
  {Gorgone}, {Kaneko}, {Baring}, {Guiriec}, {G{\"o}{\v{g}}{\"u}{\textcommabelow
  s}}, {Granot}, {Watts}, {Lin}, {Bhat}, {Bissaldi}, {Chaplin}, {Finger},
  {Gehrels}, {Gibby}, {Giles}, {Goldstein}, {Gruber}, {Harding}, {Kaper}, {von
  Kienlin}, {van der Klis}, {McBreen}, {Mcenery}, {Meegan}, {Paciesas},
  {Pe'er}, {Preece}, {Ramirez-Ruiz}, {Rau}, {Wachter}, {Wilson-Hodge}, {Woods},
  \& {Wijers}}]{Horst2012}
{van der Horst}, A.~J., {Kouveliotou}, C., {Gorgone}, N.~M., {et~al.} 2012,
  \apj, 749, 122, \dodoi{10.1088/0004-637X/749/2/122}

\bibitem[{{van Hoven} \& {Levin}(2011)}]{Hoven2011}
{van Hoven}, M., \& {Levin}, Y. 2011, \mnras, 410, 1036,
  \dodoi{10.1111/j.1365-2966.2010.17499.x}

\bibitem[{{van Hoven} \& {Levin}(2012)}]{Hoven2012}
---. 2012, \mnras, 420, 3035, \dodoi{10.1111/j.1365-2966.2011.20177.x}

\bibitem[{{Wadiasingh} {et~al.}(2020){Wadiasingh}, {Beniamini}, {Timokhin},
  {Baring}, {van der Horst}, {Harding}, \& {Kazanas}}]{Wadiasingh2020}
{Wadiasingh}, Z., {Beniamini}, P., {Timokhin}, A., {et~al.} 2020, \apj, 891,
  82, \dodoi{10.3847/1538-4357/ab6d69}

\bibitem[{{Wadiasingh} \& {Timokhin}(2019)}]{Wadiasingh2019}
{Wadiasingh}, Z., \& {Timokhin}, A. 2019, \apj, 879, 4,
  \dodoi{10.3847/1538-4357/ab2240}

\bibitem[{{Wang} {et~al.}(2020){Wang}, {Wang}, {Yang}, {Yu}, {Zuo}, \&
  {Dai}}]{WangFY2020}
{Wang}, F.~Y., {Wang}, Y.~Y., {Yang}, Y.-P., {et~al.} 2020, \apj, 891, 72,
  \dodoi{10.3847/1538-4357/ab74d0}

\bibitem[{{Wang} \& {Lai}(2020)}]{Wang2020}
{Wang}, J.-S., \& {Lai}, D. 2020, \apj, 892, 135,
  \dodoi{10.3847/1538-4357/ab7dbf}

\bibitem[{{Wang} {et~al.}(2018){Wang}, {Peng}, {Wu}, \& {Dai}}]{Wang2018}
{Wang}, J.-S., {Peng}, F.-K., {Wu}, K., \& {Dai}, Z.-G. 2018, \apj, 868, 19,
  \dodoi{10.3847/1538-4357/aae531}

\bibitem[{{Wang} {et~al.}(2016){Wang}, {Yang}, {Wu}, {Dai}, \&
  {Wang}}]{Wang2016}
{Wang}, J.-S., {Yang}, Y.-P., {Wu}, X.-F., {Dai}, Z.-G., \& {Wang}, F.-Y. 2016,
  \apj, 822, L7, \dodoi{10.3847/2041-8205/822/1/L7}

\bibitem[{{Watts} \& {Strohmayer}(2007)}]{Watts2007}
{Watts}, A.~L., \& {Strohmayer}, T.~E. 2007, Advances in Space Research, 40,
  1446, \dodoi{10.1016/j.asr.2006.12.021}

\bibitem[{{Waxman}(2017)}]{Waxman2017}
{Waxman}, E. 2017, \apj, 842, 34, \dodoi{10.3847/1538-4357/aa713e}

\bibitem[{{Yang} \& {Zhang}(2018)}]{Yang2018}
{Yang}, Y.-P., \& {Zhang}, B. 2018, \apj, 868, 31,
  \dodoi{10.3847/1538-4357/aae685}

\bibitem[{{Yang} {et~al.}(2020){Yang}, {Zhu}, {Zhang}, \& {Wu}}]{Yang2020}
{Yang}, Y.-P., {Zhu}, J.-P., {Zhang}, B., \& {Wu}, X.-F. 2020, arXiv e-prints,
  arXiv:2006.03270.
\newblock \doarXiv{2006.03270}

\bibitem[{{Yu} {et~al.}(2020){Yu}, {Zou}, {Dai}, \& {Yu}}]{Yu2020}
{Yu}, Y.-W., {Zou}, Y.-C., {Dai}, Z.-G., \& {Yu}, W.-F. 2020, arXiv e-prints,
  arXiv:2006.00484.
\newblock \doarXiv{2006.00484}

\bibitem[{{Yuan} {et~al.}(2020){Yuan}, {Beloborodov}, {Chen}, \&
  {Levin}}]{Yuan2020}
{Yuan}, Y., {Beloborodov}, A.~M., {Chen}, A.~Y., \& {Levin}, Y. 2020, arXiv
  e-prints, arXiv:2006.04649.
\newblock \doarXiv{2006.04649}

\bibitem[{{Zhang}(2016)}]{Zhang2016}
{Zhang}, B. 2016, \apj, 827, L31, \dodoi{10.3847/2041-8205/827/2/L31}

\bibitem[{{Zhang}(2020)}]{Zhang2020}
---. 2020, \apjl, 890, L24, \dodoi{10.3847/2041-8213/ab7244}

\bibitem[{Zhang {et~al.}(2020)Zhang, Jiang, Meng, Xu, Niu, \&
  et~al.}]{FAST2020}
Zhang, C.~F., Jiang, J.~C., Meng, Y. P.and~Wang, J.~B., {et~al.} 2020, The
  Astronomer's Telegram, 13699, 1

\bibitem[{{Zhong} {et~al.}(2020){Zhong}, {Dai}, {Zhang}, \& {Deng}}]{Zhong2020}
{Zhong}, S.~Q., {Dai}, Z.~G., {Zhang}, H.~M., \& {Deng}, C.~M. 2020, arXiv
  e-prints, arXiv:2005.11109.
\newblock \doarXiv{2005.11109}

\bibitem[{{Zhou} {et~al.}(2020){Zhou}, {Zhou}, {Chen}, {Wang}, {Vink}, \&
  {Wang}}]{Zhou2020}
{Zhou}, P., {Zhou}, X., {Chen}, Y., {et~al.} 2020, arXiv e-prints,
  arXiv:2005.03517.
\newblock \doarXiv{2005.03517}

\end{thebibliography}

\end{CJK*}
\end{document}